\documentclass[journal]{IEEEtran}
\ifCLASSINFOpdf
 
\else
\usepackage[dvips]{graphicx}
\fi
\usepackage[cmex10]{amsmath}

\usepackage{amsmath,amsthm,amsfonts,amssymb,latexsym,extarrows,booktabs,array}
\usepackage[cmex10]{amsmath}
\usepackage{subfigure}
\usepackage{threeparttable}
\usepackage{mathrsfs}
\usepackage{multicol}
\usepackage{amsmath,amsthm,amsfonts,amssymb,latexsym,extarrows,booktabs,array}
\usepackage{stfloats}
\usepackage{algorithm}
\usepackage{algorithmic}
\newcommand{\ud}{\mathop{}\!\mathrm{d}}

\begin{document}

\title{On the Deployment of Distributed Antennas for Wireless Power Transfer with Safety Electromagnetic Radiation Level  Requirement}

\author{Chao Zhang, and Guanghe Zhao
\thanks{C. Zhang and G. Zhao are with School of Electronics and Information Engineering, Xi'an Jiaotong University, Xi'an, 710049 China. E-mail: \{chaozhang@mail.xjtu.edu.cn, zhaoguanghe2010@stu.xjtu.edu.cn\}.}
\thanks{This work was supported in part by Natural Science Basic Research Plan in Shaanxi Province of China (Program No.2015JQ6234), the National Natural Science Foundation of China (NSFC) under Grant (No.61461136001), and the Fundamental Research Funds for the Central Universities.}}

\maketitle
\begin{abstract}
The extremely low efficiency is regarded as the bottleneck of Wireless Power Transfer (WPT) technology.  To tackle this problem, either enlarging the transfer power or changing the infrastructure of WPT system could be an intuitively proposed way. However, the  drastically important issue on the user exposure of electromagnetic radiation is rarely considered while we try to improve the efficiency of WPT.  In this paper, a Distributed Antenna Power Beacon (DA-PB) based WPT system where these antennas are uniformly distributed on a circle is analyzed and optimized with the safety electromagnetic radiation level (SERL) requirement. In this model, three key questions are intended to be answered: 1) With the SERL, what is the performance of the harvested power at the users ?  2) How do we configure the parameters to maximize the efficiency of WPT?  3) Under the same constraints, does the DA-PB still have performance gain than the Co-located Antenna PB (CA-PB)?  First,  the minimum antenna height of DA-PB is derived to make the radio frequency (RF) electromagnetic radiation power density at any location of the charging cell lower than the SERL published by the  Federal Communications Commission (FCC). Second, the closed-form expressions of average harvested Direct Current (DC) power per user in the charging cell for pass-loss exponent 2 and 4 are also provided. In order to maximize the average efficiency of WPT, the optimal radii for distributed antennas elements (DAEs) are derived when the pass-loss exponent takes the typical value $2$ and $4$. For comparison, the CA-PB is also analyzed as a benchmark. Simulation results verify our derived theoretical results. And it is shown that the proposed DA-PB indeed achieves larger average harvested DC power than CA-PB and can improve the efficiency of WPT. 
\end{abstract}


\section{Introduction}
\IEEEpeerreviewmaketitle
\IEEEPARstart{D}{espite} of the significant advances in Wireless Power Transfer (WPT), there are a lot of open issues that are summarized as follows: First, the transfer distance in WPT is stringently limited and desperately need to be increased. It is known that the signal power attenuates by the exponent of transfer distance. In order to get viable received power, the distance is generally severely small thus restricts its application in electronics such as portable and wearable electronics. Second, wireless power transfer efficiency, which is becoming a vital metric, is extremely small based on the current state-of-the-art research and also needs to be improved.

\subsection{Context and Motivation}
Wireless power transfer (WPT) has recently drawn more and more attention due to that it enables proactive energy replenishment of user terminals. There are two related research topics, i.e., simultaneous wireless information and power transfer (SWIPT) and PB-assisted wirelessly powered communication networks (PB-assisted WPCN). The study of SWIPT can be referred to \cite{Survey1}-\cite{Survey4} and references therein. Compared with the point-to-point SWIPT, the authors in \cite{Survey5} proposed an iterative dynamic power splitting algorithm to maximize the receiving signal-to-noise ratio (SNR) at the destination node for the multi-relay networks with wireless energy harvesting. SWIPT is suitable for the case where users are close to the base station (BS). It is due to the fact that the operating power of the energy harvesting component is generally much higher than that of the information decoding component \cite{Survey6}. Compared with SWIPT, PB-assisted WPCN system generally has a larger coverage region. Furthermore, the users in PB-assisted WPCN tend to harvest more energy. \par
The other research topic focuses on the PB-assisted WPCN. Three different configurations for a wireless-powered cellular network were investigated in \cite{Survey7}. The first was full-duplex BS with energy transfer in the downlink and information transfer in the uplink; In the second configuration, distributed PBs were exploited to power the user nodes and the power harvested at the user was used to transmit information to the BS; In the third configuration, distributed PBs and distributed antenna elements (DAEs) were considered. The authors argued that by exploiting distributed PBs, the system performance could be significantly improved. However, \cite{Survey7} did not consider the RF electromagnetic radiation, which is extremely indispensable and draws more and more attention in practice. In \cite{Survey8}, the authors proposed a novel multi-user scheduling strategy, i.e., opportunistic scheduling, and analyzed its performance gain in two systems namely homogeneous and heterogeneous users system over the round-robin scheduling. It is worthy to point that the safety radiation was considered in \cite{Survey8}. The authors in \cite{Survey9} proposed an adaptively directional WPT scheme for power beacon to improve the efficiency in a large WPT system. Specifically, the power beacon can adaptively perform energy beamforming according to the number of users and their locations in order to lead the power to the users within the charging region of power beacon. Unfortunately, the authors in \cite{Survey9} did not consider the electromagnetic exposure either. \par
As a mature technology, Distributed Antenna Systems (DAS) has been shown to have the ability to significantly increase coverage as well as improve system throughput \cite{Survey7}, \cite{Survey10}-\cite{Survey13}. Uniform circular layout (UCL) of DAEs was generally exploited to analyze the performance of DAS in company with circular cell \cite{Survey7}, \cite{Survey10}-\cite{Survey13}. In this paper, we pursue the work of DAS and investigate the optimal deployment of PB DAEs with uniform circular layout.\par
\subsection{Contributions and Organization of the Paper}
The contributions of this paper are summarized as follows:
\begin{itemize}
    \item A novel deployment architecture of antennas for PB is proposed to implement efficient WPT. Considering the radio frequency (RF) electromagnetic radiation safety level drafted by the Federal Communications Commission (FCC), we get the closed-form expression of DA-PB antenna height to make the RF electromagnetic radiation power density at any location of the charging cell lower than the safety level limited by FCC.
    \item For the proposed DA-PB, we give the closed-form result of average harvested DC power per user in the charging cell when path-loss exponent takes the typical value $2$ and $4$, which are the typical values for suburban area and urban city, respectively.
    \item In order to maximize the average efficiency of wireless power transfer, we get the optimal radii of distributed antennas of DA-PB when path-loss exponent takes the typical value $2$ and $4$.
\end{itemize} \par
The remainder of the paper is organized as follows. Section \ref{sec:system_model} elaborates the system model. The calculation of antenna height of DA-PB and the performance analysis are presented in Section \ref{sec:calculation_ave_DC_P}. In Section \ref{sec:location_opti_DA}, in order to maximize the average efficiency of WPT when using DA-PB, we get the optimal radii of distributed antennas when path-loss exponent takes the typical value $2$ and $4$. Simulation results and discussion are presented in Section \ref{sec:simulations_discussion}. Finally, Section \ref{sec:conclusion} concludes the paper and followed by detailed derivation process of some results relegated to appendices.\par
\textbf{Notation:} For a complex variable $x$, operators $\Re\{x\}$, $|x|$ and $\arg(x)$ denote its real part, amplitude and phase, respectively. $\mathbb{E}_y\{x\}$ stands for the statistical expectation of real random variable $x$ with respect to $y$ and $x\sim \mathcal{U}(a,b)$ denotes that $x$ is a random variable following the uniform distribution in the interval from $a$ to $b$. Finally, $\overline{P}_{out-x}$ stands for the average harvested DC power per user, where $x \in \{\mathrm{CA, DA}\}$ stands for the deployment structure of the PB antennas ('CA' for co-located antennas and 'DA' for distributed antennas). $\overline{\eta}_{x}$ stands for the average efficiency of WPT, where the meaning of $x$ is similar to that in $\overline{P}_{out-x}$.
\section{System Model}\label{sec:system_model}

\begin{figure*}[ht]
\centering
\subfigure[Conventional CA-PB with multi-antennas]{
\label{model_1a}
\includegraphics[width=8cm]{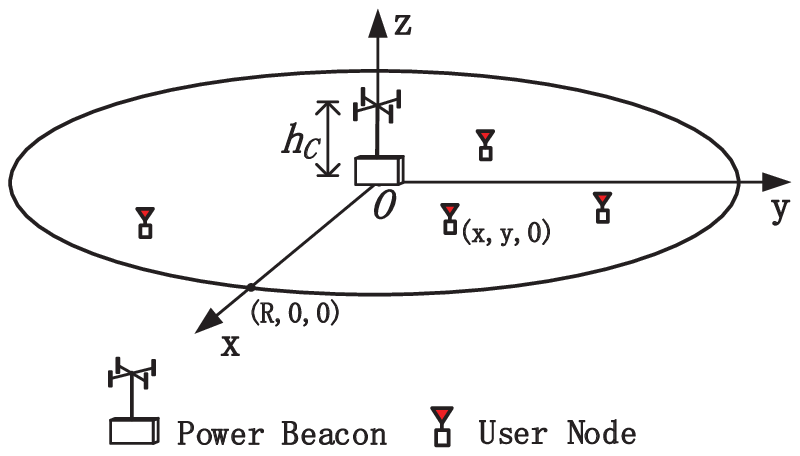}}
\subfigure[Our proposed DA-PB with multi-antennas]{
\label{model_1b}
\includegraphics[width=8cm]{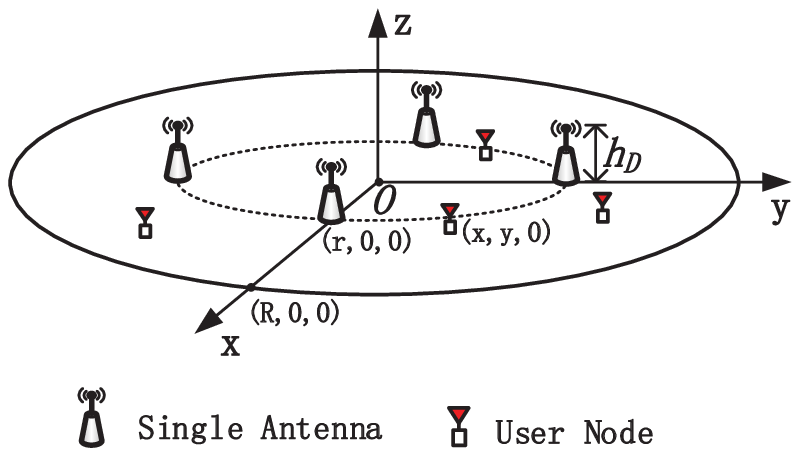}}
\caption{System model.}
\label{fig_system_model}
\end{figure*}

As depicted in Fig.\ref{fig_system_model}, we assume that the region covered by the PB is a circle with the radius $R$. Suppose the PB has $N$ antennas with the total power $P$ and each user has a single antenna. For the convenience of illustration, $N$ equals to $4$ in Fig.\ref{fig_system_model}. The users whose height is assumed to be zero are uniformly distributed in the charging cell. Specifically, in Fig.\ref{model_1a}, the PB with multi-antennas is located at the center of the circle and the distances between those antennas are extremely small compared to the distances from the PB to the users. Thus it can be deemed as the so-called Co-located Antenna Power Beacon (CA-PB). We denote the antenna height of CA-PB as $h_C$.
In contrast, the PB in Fig.\ref{model_1b} is the Distributed Antenna Power Beacon (DA-PB). The distributed antenna elements (DAEs) of DA-PB are uniformly deployed on the circle whose radius is $r$ and might be connected to a central power source through power lines or different power sources. We further assume that the PB has no knowledge of the channel state information (CSI) between the PB and the users, so equal power allocation among the antennas is considered in this paper, i.e., the transmit power of each antenna is $P/N$. Suppose all antennas of DA-PB have the same height $h_D$. To let the radiation power density at any location of the charging cell  lower than the SERL published by FCC, we should set $h_C$ and $h_D$ carefully.  
\subsection{Signal Propagation Model}
The power transmitted by each antenna of PB can be aggregated at the user. The RF signal transmitted by the $i^{\rm th}$ antenna at time slot $t$ can be expressed as 
\begin{equation}
{s}_{i}(t)=\sqrt{2P_i}\Re\left\{x_i(t)e^{j2\pi ft}\right\}, 
\end{equation}
where $P_i$ denotes the transmit power of the $i^{\rm th}$ antenna, $f$ refers to the carrier frequency, and $x_i(t)$ is the complex baseband signal with bandwidth $B$ Hz and unit power, i.e., $|x_i(t)|^2=1$. It is assumed that $B \ll f$. For a fixed user, the received signal at the user is
\begin{equation}
{r(t)}=\sum\limits_{i=1}^N\sqrt{\frac{2P_i c|h_i(t)|^2}{d_i^\alpha}}\Re \left\{x_i(t)e^{j[2\pi ft+\theta_i(t)]} \right\}+n(t),
\end{equation}
where $c$ stands for the constant scaling factor, $d_i$, $\theta_i(t)$, and $|h_i(t)|^2$ denote the distance, phase shift, and power gain of the fast fading channel from the $i^{\rm th}$ antenna to the user, respectively. Additionally, $n(t)$ is the additive white Gaussian noise (AWGN) at the user at time slot $t$. Compared to the received RF signal, the noise power is usually greatly small thus can be neglected. Therefore
\begin{equation}
\begin{split}
{r(t)} &\approx \sum\limits_{i=1}^N\sqrt{\frac{2P_i c|h_i(t)|^2}{d_i^\alpha}}\Re \left\{x_i(t)e^{j[2\pi ft+\theta_i(t)]} \right\} \\
    & = \sum\limits_{i=1}^N\sqrt{\frac{2P_i c|h_i(t)|^2}{d_i^\alpha}}\cos \left[2\pi ft+\arg\left(x_i(t)\right)+\theta_i(t) \right].
\end{split}
\end{equation}

At the energy receiver of the user, the received RF signal first goes through the nonlinear Schottky diode, thus the output current includes the DC component as well as the harmonic components at $kf ~(k\geq1)$. Due to the Shockley's diode equation \cite{Survey14}, the output current after the Schottky diode at time slot $t$ is 
\begin{equation}\label{i_t}
\begin{split}
&{i(t)} \!=\! I_s \left(e^\frac{r(t)}{\rho V_T}\!-\!1 \right)\!=\!\sum\limits_{k=1}^{\infty}\frac{I_s}{k!(\rho V_T)^k}r^k(t) \!\approx\! \frac{I_s}{2(\rho V_T)^2}r^2(t)  \\
\end{split}
\end{equation}
where $I_s$ denotes the reverse saturation current of the diode, $\rho$ is the ideality factor of the diode \footnote{ The ideality factor of the diode $\rho$ generally has a range between $1$ and $2$, which depends upon the operating conditions and physical construction.}, and $V_T$ refers to the thermal voltage. The second equation in (\ref{i_t}) is derived by exploiting Taylor series expansion of the exponential function. After rectifying, we only consider the quadratic term of output signal because the coefficients of the higher-order $(k>2)$ terms in (\ref{i_t}) is very small  \cite{Survey4}\cite{Survey8}.  After that, the output current $i(t)$ is fed into the low pass filter (LPF). Then the direct current signal without high frequency components is
\begin{equation}\label{i_dc_t_temp}
\begin{split}
&{i}_{\rm dc}(t) =\frac{I_sc}{2(\rho V_T)^2}\left\{\sum\limits_{i=1}^N \frac{P_i |h_i(t)|^2}{d_i^\alpha} \right.\\
    & +\sum\limits_{i=1}^N\sum\limits_{j=1,j\neq i}^N \sqrt{\frac{P_i P_j |h_i(t)|^2 |h_j(t)|^2}{d_i^\alpha d_j^\alpha}} \\
    & \times \left[ \cos\left[\arg\left(x_i(t)\right)+\theta_i(t)\right]\cos\left[\arg\left(x_j(t)\right)+\theta_j(t)\right] \right.\\
    & \left. \left. + \sin\left[\arg\left(x_i(t)\right)+\theta_i(t)\right]\sin\left[\arg\left(x_j(t)\right)+\theta_j(t)\right] \right] \right\}
\end{split}, 
\end{equation}
in which $\theta_i(t)$ and $|h_i(t)|^2$ are the phase and the power gain of fast Rayleigh fading channel, respectively.  $\theta_i(t)$ is an uniformly distributed variable, i.e., $\theta_i(t) \sim \mathcal{U}(-\pi,\pi)$ and $|h_i(t)|^2$ is a random variable following the exponent distribution~\cite{Survey15}. In addition,  $\{\theta_1(t), \theta_2(t),...,\theta_N(t) \}$ and $\{|h_1(t)|^2, |h_2(t)|^2,...,|h_N(t)|^2\}$ are independently identically distributed (i.i.d.),  respectively. Note that $\theta_i(t)$ and $|h_i(t)|^2$ are   independent on $d_i$ and $x_i(t)$. The probability density function (PDF) of  $|h_i(t)|^2$ is
\begin{equation}
\begin{split}
&{f}_{|h_i(t)|^2}(\zeta)=
\begin{cases}
\frac{1}{\sigma_h^2}e^{-\frac{\zeta}{\sigma_h^2}}, & \mbox{if}~\zeta>0,\\
0, & \mbox{otherwise}.\\
\end{cases}
\end{split}
\end{equation}
where $\sigma_h^2$ denotes the mean of the random variable $|h_i(t)|^2$. After averaging the random phase $\theta_i(t)$ and $|h_i(t)|^2$, we get the average DC current as 
\begin{equation}
\begin{split}
\overline{i}_{\rm dc}(t) \!=\! \frac{I_s c}{2(\rho V_T)^2} \! \sum\limits_{i\!=\!1}^N \! \mathbb{E}_{|h_i(t)|^2} \! \left\{\frac{P_i |h_i(t)|^2}{d_i^\alpha} \! \right\} \!=\! \frac{I_s c \sigma_h^2}{2(\rho V_T)^2} \! \sum\limits_{i=1}^N \! \frac{P_i }{d_i^\alpha}
\end{split}
\end{equation}
Finally, the DC current is converted to the DC power and then stored in the rechargeable battery. The power charged to the battery is generally linearly proportional to the input DC with the scaling factor being energy transfer efficiency $0<\xi < 1$. Thus the ergodic harvested DC power $\overline{P}_{\rm out}(x,y,0,t)$ for the user at the coordinate of $(x,y,0)$ is given by
\begin{equation}\label{ergodic_harvested_DC_power}
\overline{P}_{\rm out}(x,y,0,t) =\xi \overline{i}_{\rm dc}(t)=K_0 \sum\limits_{i=1}^N \frac{P_i}{d_i^\alpha}
\end{equation}
where $K_0\triangleq\frac{\xi I_s c \sigma_h^2}{2(\rho V_T)^2}$ is a constant. Note that (\ref{ergodic_harvested_DC_power}) is actually the sum of average received power transmitted from different antennas. Thus we have completed the proof of our assumption. It is worth mentioning that (\ref{ergodic_harvested_DC_power}) is similar to those in \cite{Survey7}\cite{Survey9}\cite{Survey16}, which verifies our assumption and derivation. In addition, we assume that a quasi-static block-fading is considered and the channel gain from the antenna to the user is independent from block to block. Therefore, for the convenience of illustration, we discard the index $t$ in the remainder of the paper. \par
\newtheorem{remark}{Remark}
\begin{remark}\label{remark_phase_opt}{(Technology of Maximizing Instantaneous Harvested DC Power):}
\emph{We admit that by elaborately designing the power allocation and transmission phase in (\ref{i_dc_t_temp}), the instantaneous harvested DC power of a user can be maximized (see \cite{Survey17} and references therein), but this will need estimation and feedback of the instantaneous CSI. First, the estimation and feedback are generally not as accurate as expected, which hinders us from getting optimal system performance and even deteriorates the system performance; Second, estimation and feedback of CSI increase the system overhead. Thus, in this paper, we consider the ergodic harvested DC power of (\ref{i_dc_t_temp}). Note that the DA-PB without any extra estimation and feedback of CSI discussed in this paper is quite easy to implement in practice.}
\end{remark}

\subsection{Radio Frequency Electromagnetic Radiation}
Considering the safety levels of human exposure to RF electromagnetic fields, we place the antennas at the height of $h_C$ and $h_D$ for CA-PB and DA-PB, respectively. Generally speaking, because the industrial, scientific, and medical (ISM) frequency band is open and free, WPT can use the ISM band such as 2.45 GHz, 5.8 GHz to perform WPT in practice \cite{Survey8}. The radiation power density is computed via $\Psi=\frac{P_r}{4\pi d^2}$ (see \cite{Survey18}, p. 32) where $\Psi$, $P_r$, $d$ are the radiation power density at the distance of $d$ from the power beacon, power beacon transmit power, and the distance between the user and the power beacon, respectively. \par

\section{Antenna Height of PB and Performance Analysis}\label{sec:calculation_ave_DC_P}
In this section, considering the equal power allocation among antennas, we first derive the minimum antenna height of PB in order to protect users from being hurt by RF electromagnetic radiation. Then, we analyze the average harvested DC power per user in the charging cell and the average efficiency of WPT for CA-PB and our proposed DA-PB. In addition, the users follow the uniform distribution in the charging cell. The system performance is characterized by the average harvested DC power per user and the average efficiency of WPT. Specifically, we average the resultant ergodic harvested DC power in the whole cell and yield the average harvested DC power per user. The average efficiency of WPT, which can be exploited to judge what kind of deployment is more energy efficient, is defined as the ratio of average harvested DC power per user and the total transmit power of PB.
\subsection{Antenna Height of PB}
\subsubsection{CA-PB}
For CA-PB, the transmit power $P$ and the antenna height $h_C$ of power beacon should be limited by (\ref{safety_level_equation}) in order to avoid exclusion zone in the charging cell (We are only interested in the disc whose height is zero because users height is assumed to be zero). By the way, to avoid exclusion zone is referred to making the radiation power density at any location in the charging cell lower than the safety level.
\begin{equation} \label{safety_level_equation}
P < 4\pi h_C^2\Psi_0
\end{equation}
where $\Psi_0$ denotes the safety radiation level\footnote{ According to the IEEE standard C95.1-2005, the safety radiation level of human exposure to RF electromagnetic fields from 2 GHz to 100 GHz is 10 $\mathrm{W/m^2}$ (i.e., 1 $\mathrm{mW/cm^2}$) (\cite{Survey8} and \cite{Survey19}, p. 27).} given by FCC. This result can offer useful directions when deploying the CA-PB antennas to avoid exclusion zone in the charging cell.\par
\subsubsection{DA-PB}
For DA-PB with uniform circular layout (UCL) of DAEs and equal power allocation (see Fig.\ref{model_1b}), without loss of generality, we assume the coordinates of the DAEs are listed as follows. For the convenience of expression, we have assigned a number for each DAE. Specifically, we denote the coordinate of ${\rm DAE}_i$ as
\begin{equation}\label{DAE_i_location}
\begin{split}
&O_i\!=\!\left(r\cos\frac{2\pi(i-1)}{N},r\sin\frac{2\pi(i-1)}{N},h_D\right),
\end{split}
\end{equation}
We aim to derive the expression of $h_D$ in the remainder of this subsection with which to avoid exclusion zone in the charging cell. In other words, the maximal radiation density in the charging cell must be lower than the safety level. For DA-PB with DAEs located as (\ref{DAE_i_location}), because of the symmetry property, the coordinates of maximal radiation density in the disc must be as follows
\begin{equation}
E_i\!=\!\left(\nu^\star \cos\frac{2\pi(i-1)}{N},\nu^\star \sin\frac{2\pi(i-1)}{N},0\right),  \\
\end{equation}
where $\nu^\star$ is the distance between the maximal radiation density coordinates and the charging cell center. All the radiation power densities at each $E_i$ are the same, so we only consider the first coordinate of maximal radiation density without loss of generality. The equality of maximal electromagnetic radiation density in the charging cell for CA-PB and DA-PB guarantees the fairness of CA-PB and DA-PB in order to compare their performance. In addition, suppose the maximal radiation density is lower than the safety level given by FCC. So we straightly get
\begin{equation} \label{radiation_CA_equal_to_DA}
\begin{split}
&\frac{P}{4\pi h_C^2}\!=\!\sum\limits_{i=1}^N \frac{\frac{P}{4\pi N}}{\left(\nu^\star\!-\!r\cos\frac{2\pi(i-1)}{N}\right)^2\!+\!\left(r\sin\frac{2\pi(i-1)}{N}\right)^2\!+\!h_D^2}
\end{split}
\end{equation}
It is hard to give a closed-form expression of $\nu^\star$ and $h_D$ from (\ref{radiation_CA_equal_to_DA}). However, when $N\rightarrow\infty$, we get explicit simple analytical expressions for $\nu^\star$ and $h_D$, i.e.,
\begin{equation}
\begin{split}
&{\nu^\star}=
\begin{cases}
0, & \mbox~0\leq r \leq \frac{h_C}{\sqrt{2}},\\
\sqrt{r^2-\left(\frac{h_C^2}{2r}\right)^2}, & \mbox ~\frac{h_C}{\sqrt{2}}\leq r \leq R.\\
\end{cases}
\end{split}
\end{equation}
and
\begin{equation}\label{h_D_vs_r}
\begin{split}
&{h_D}=
\begin{cases}
\sqrt{h_C^2-r^2}, & \mbox~0\leq r \leq \frac{h_C}{\sqrt{2}},\\
\frac{h_C^2}{2r}, & \mbox ~\frac{h_C}{\sqrt{2}}\leq r \leq R.\\
\end{cases}
\end{split}
\end{equation}
The detailed derivation process can be found in Appendix A. Note that $\nu^\star$ is piecewise function of DAE radius $r$, to speak specifically, a non-decreasing function, and is continuous at the point of $r=\frac{h_C}{\sqrt{2}}$. However, $h_D$ is a decreasing function of $r$, and is also continuous at the point of $r=\frac{h_C}{\sqrt{2}}$.
\subsection{Average Harvested DC Power}
\subsubsection{CA-PB}
Compared with the distance between the antennas and the user, the distance between antennas in CA-PB is extremely smaller, so we regard all the PB antennas as co-located so as to simplify the analysis. Thus the distance between different antennas and the user is the same. Without loss of generality, we assume the coordinate of the user is $(x,y,0)$. The distance between the CA-PB antennas and the user is denoted as $d_0=\sqrt{x^2+y^2+h_C^2}$. By virtue of (\ref{ergodic_harvested_DC_power}), the ergodic harvested DC power of the user at $(x,y,0)$ is
\begin{equation}\label{ave_P_out_CAS_x_y}
\overline{P}_{\rm out-CA}(x,y,0)=K_0 \frac{P}{d_0^\alpha}
\end{equation}
Assume that the users are uniformly distributed in the charging cell, we thereby straightly give the probability density function (PDF) when user node is at the coordinate of $(x,y,0)$
\begin{equation}\label{pdf_x_y}
\begin{split}
&{f}(x,y,0)=
\begin{cases}
\frac{1}{\pi R^2}, & \mbox{if}~x^2+y^2\leq R^2,\\
0, & \mbox{otherwise}.\\
\end{cases}
\end{split}
\end{equation}
So the average harvested DC power per user in the charging cell is
\begin{equation}\label{ave_P_out_CAS}
\overline{P}_{\rm out-CA}=K_0\frac{2P}{(\alpha-2) R^2}\left[\frac{1}{h_C^{\alpha-2}}-\frac{1}{(R^2+h_C^2)^{\frac{\alpha}{2}-1}}\right]
\end{equation}
For the special case, when $\alpha$ takes the value $2$, we get
\begin{equation}
   \overline{P}_{\rm out-CA}=\frac{K_0 P}{R^2}\ln\left(1+\frac{R^2}{h_C^2}\right)
\end{equation}
It is obvious that the average harvested DC power per user for CA-PB linearly increases as the transmit power goes up.

\subsubsection{DA-PB}
Compared with CA-PB, the distances between the DAEs and user in the DA-PB are usually different. ${d}_{i}\!=\!\sqrt{\left(x\!-\!r\cos\frac{2\pi(i\!-\!1)}{N}\right)^2\!+\!\left(y\!-\!r\sin\frac{2\pi(i\!-\!1)}{N}\right)^2\!+\!h_D^2},\forall i\in[1,N]$ denotes the distance between ${\rm DAE}_i$ and the user. Consequently, the ergodic harvested power of the user at $(x,y,0)$ is
\begin{equation}\label{ave_P_out_DAS_All_x_y}
\overline{P}_{\rm out-DA}(x,y,0)=K_0\frac{P}{N}\sum\limits_{i=1}^{N}\frac{1}{d_i^\alpha}
\end{equation}
By averaging (\ref{ave_P_out_DAS_All_x_y}), we get
\begin{equation}\label{ave_P_out_DAS_All}
\begin{split}
\overline{P}_{\rm out-DA}&\!=\!\int_{0}^{2\pi}\! \! \int_{0}^{R} \! \frac{K_0 P}{N\pi R^2}\!\sum\limits_{i=1}^{N}\!\frac{1}{d_i^{\alpha}}\rho \ud\rho \ud\theta \\
&\!=\! \frac{K_0 P}{\pi R^2}\underbrace{\int_{0}^{2\pi}\!\int_{0}^{R}\!\frac{\rho}{d_1^\alpha}\ud\rho \ud\theta}_{Q}
\end{split}
\end{equation}
in which the symmetry property has been used to get (\ref{ave_P_out_DAS_All}). $Q$ is intractable but we get an explicit closed-form expression when $\alpha$ takes the typical value $2$ and $4$ as follows
\begin{equation}\label{Q_alpha_2_4}
\begin{split}
&{Q}\!=\!
\begin{cases}
\pi \ln\left(\frac{R^2+h_D^2-r^2+\sqrt{(R^2+h_D^2-r^2)^2+4r^2h_D^2}}{2h_D^2}\right), & \mbox~\alpha=2,\\
\pi \frac{R^2-h_D^2-r^2+\sqrt{R^4+R^2(2h_D^2-2r^2)+(r^2+h_D^2)^2}}{2h_D^2 \sqrt{R^4+R^2(2h_D^2-2r^2)+(r^2+h_D^2)^2}}, & \mbox ~\alpha=4.\\
\end{cases}
\end{split}
\end{equation}
The detailed derivation process of (\ref{Q_alpha_2_4}) is presented in Appendix B. Note that the $\overline{P}_{\rm out-DA}$ is also proportional to the transmit power because the definite integral $Q$ in (\ref{ave_P_out_DAS_All}) is actually a constant and  power independent. \par
On the other hand, let $r_{MS}$ denote the distance between the user and the cell center, then
\begin{equation}
\begin{split}
&\overline{P}_{\rm out-DA}\!(r_{MS}\!) \!=\! \!\sum\limits_{i=1}^N\! \frac{K_0 P}{N\!\left(\!r_{MS}^2\!+\!r^2\!-\!2rr_{MS}\!\cos\frac{2\pi(i\!-\!1)}{N}\!\!+\!h_D^2\!\right)^\frac{\alpha}{2}}
\end{split}
\end{equation}
When $N \rightarrow \infty$, we get
\begin{equation}\label{P_out_DA_r_MS}
\begin{split}
&\lim_{N\rightarrow\infty}{\overline{P}}_{\rm out-DA}(r_{MS}) \\
&= \!\frac{K_0 P}{N} \!\frac{N}{2\pi}\!\sum\limits_{i=1}^N\! \frac{1}{\left(r_{MS}^2\!+\!r^2\!-\!2rr_{MS} \cos\frac{2\pi(i-1)}{N}\!+\!h_D^2\right)^\frac{\alpha}{2}}\! \frac{2\pi}{N} \\
&=\frac{K_0 P}{2\pi} \int_{0}^{2\pi} \frac{1}{\left(r_{MS}^2+r^2-2rr_{MS} \cos\theta+h_D^2\right)^\frac{\alpha}{2}} \ud\theta \\
&= K_0 P \left[(r_{MS}^2+r^2+h_D^2)^2-4r^2r_{MS}^2\right]^{-\frac{\alpha}{4}} \\
& ~~~~\times P_{\frac{\alpha}{2}-1}\left(\frac{r_{MS}^2+r^2+h_D^2}{\sqrt{(r_{MS}^2+r^2+h_D^2)^2-4r^2r_{MS}^2}}\right) \\
\end{split}
\end{equation}
where $P_\cdot (\cdot)$ denotes the Legendre function (\cite{Survey20}) and $P_a(b)=F(-a,a+1;1;\frac{1-b}{2})$, where $F(\cdot,\cdot;\cdot;\cdot)$ is the Gauss hypergeometric function (\cite{Survey20}). This function can be calculated by using any standard mathematical software packages such as MATLAB and MAPLE. Note that we have used (\cite{Survey21}, (2.5.16.38)) to get the last equation in (\ref{P_out_DA_r_MS}).
\subsection{Average Efficiency of WPT}
In our system, the average efficiency of WPT is defined as the ratio of average harvested DC power per user and the PB transmit power. The average efficiency of WPT can be deemed as an extraordinarily important metric when judging which deployment of antennas for PB is more energy efficient.
\subsubsection{CA-PB}
Note that all the antennas simulcast energy signal to the user, thus the total transmit power is $P$. The average efficiency of WPT for CA-PB is
\begin{equation}\label{ave_eff_WPT_CAS}
\overline{\eta}_{\rm CA} \triangleq \frac{\overline{P}_{\rm out-CA}}{P} = \frac{2K_0 }{(\alpha-2)R^2} \left[ \frac{1}{h_C^{\alpha-2}}-\frac{1}{(R^2+h_C^2)^{\frac{\alpha}{2}-1}} \right]
\end{equation}
From the result above, we can argue that the average efficiency of WPT of CA-PB is determined by the antenna height of PB and the path-loss exponent.

\subsubsection{DA-PB}
Similarly, the average efficiency of DA-PB is
\begin{equation}\label{ave_eff_WPT_DAS_all}
\overline{\eta}_{\rm DA} \triangleq \frac{\overline{P}_{\rm out-DA}}{P} = \frac{K_0}{\pi R^2}Q
\end{equation}
Note that $Q$ is a variable related to path-loss exponent, antenna height of DA-PB, and the DAE radius. So we can optimize the location of DAEs to maximize the average efficiency of WPT for DA-PB.

\section{Location Optimization of Circular PB Distributed Antennas}\label{sec:location_opti_DA}
On one hand, in order to satisfy the Friis Equation, we have $h_D \geq d_{ref}$, where $d_{ref}$ is a reference distance for the antenna far field. According to (\ref{h_D_vs_r}), we get
\begin{equation}
   \frac{h_C^2}{2R} \geq d_{ref}
\end{equation}
Without loss of generality, we use $d_{ref}=1$ throughout this paper, thus $h_C \geq \sqrt{2R}$. On the other hand, in order to improve the average efficiency of WPT, the antenna height is as lower as better but it must satisfy the safety radiation level limited by FCC. Given this, we assume that $h_C < R$. Antenna height of BS being lower than the cell radius is a common assumption in the existing wireless communications related literatures. From the above, we only focus on $\sqrt{2R} \leq h_C < R$ from now on to continue our analysis.\par
\subsection{Path-Loss Exponent 2}
When $\alpha=2$, in order to maximize the average efficiency of WPT, we formulate an optimization problem to get the optimal DAE radius as follows
\begin{equation}
\begin{split}
\mathbf P1: &\max_{r }~\Upsilon_1(r)\\
&~\mathrm{s.t.}~~ 0 \leq r\leq {R}
\end{split}
\end{equation}
where $\Upsilon_1(r)={\frac{K_0}{R^2}\ln\left(\frac{R^2+h_D^2-r^2+\sqrt{(R^2+h_D^2-r^2)^2+4r^2h_D^2}}{2h_D^2}\right)}$ and $h_D$ is given by (\ref{h_D_vs_r}). We get the closed-form expression of the optimal DAE radius as follows
\begin{equation} \label{opt_r_alpha_2}
   r^\star=\frac{1}{2}\sqrt{R^2+\sqrt{R^4+4h_C^4}}
\end{equation}
The detailed derivation process can be found in Appendix C. From the startlingly concise result, we can easily find that the optimal DAE radius is only determined by the size of the cell, i.e., the radius of the cell, and the CA-PB antenna height. Note that $h_C$ is essentially determined by the safety level of radiation power density and total transmit power. This can be explained by (\ref{safety_level_equation}).\par
\subsection{Path-Loss Exponent 4}
Similarly to that when $\alpha=2$, we formulate an optimization problem to get the optimal DAE radius for $\alpha=4$ as follows
\begin{equation}
\begin{split}
\mathbf P2: &\max_{r }~\Upsilon_2(r)\\
&~\mathrm{s.t.}~~ 0 \leq r\leq {R}
\end{split}
\end{equation}
where $\Upsilon_2(r)\!=\!{\frac{K_0}{2R^2}\frac{R^2-h_D^2-r^2+\sqrt{R^4+R^2(2h_D^2-2r^2)+(r^2+h_D^2)^2}}{h_D^2 \sqrt{R^4+R^2(2h_D^2-2r^2)+(r^2+h_D^2)^2}}}$. We reformulate the above optimization problem into finding the desired real root in the range $(\frac{h_C^2}{2}, R^2)$ for the next eight-order polynomial equation
\begin{equation}\label{eight_order_equation}
p(x)=0
\end{equation}
where $p(x) = 256x^8-768R^2x^7+128(6R^4+h_C^4)x^6+(224h_C^4R^2-256R^6)x^5-192R^4h_C^4x^4-32R^2h_C^4(R^4+2h_C^4)x^3-8h_C^8(4R^4+h_C^4)x^2-10R^2h_C^{12}x-h_C^{16}$. The proof can be referred to Appendix D. It is easy to show that $p(\frac{h_C^2}{2})<0$ and $p(R^2)>0$, so there must be at least one real root for $x \in (\frac{h_C^2}{2}, R^2)$ under the condition $\sqrt{2R} \leq h_C < R$. \par
However, it is nontrivial to prove the uniqueness of real root of the above equation. We admit that we can not prove it directly. Next, we present some alternative methods to help to bracket the real roots of the above equation. Note that for $\sqrt{2R} \leq h_C < R$, only the coefficients of the eight-order and six-order terms are positive, the other coefficients are negative. According to $Descartes' ~rule ~of ~signs$ \cite{Survey22}, the number of positive real roots of the above single-variable polynomial is either equal to the number of sign differences between consecutive nonzero coefficients, or is less than it by an even number. Multiple roots of the same value are counted separately. So it is easy to argue that (\ref{eight_order_equation}) has one or three positive real roots. We further determine the number of real roots in the range $x \in (\frac{h_C^2}{2}, R^2)$ of (\ref{eight_order_equation}) by the $Sturm's ~Theorem$ \cite{Survey23}. \par
First, we get the $Sturm ~Sequence$ of $p(x)$ as: $p_0(x)=p(x), ~ p_1(x)=p'(x), ~p_2(x)=-rem(p_0,p_1)=p_1(x)q_0(x)-p_0(x), ~p_3(x)=-rem(p_1,p_2)=p_2(x)q_1(x)-p_1(x), ~\cdots, ~0=-rem(p_{m-1},p_m)$. $rem(p_i,p_j)$ and $q_i$ are the remainder and the quotient of the polynomial long division of $p_i$ by $p_j$, and $m$ is the minimal number of polynomial divisions (never greater than $deg(p)$, the degree of $p$) needed to obtain a zero remainder. Then, let $\sigma(\varsigma)$ denote the number of sign changes (ignoring zeroes) in the Sturm Sequence $ \left[ p_0(\varsigma), p_1(\varsigma), p_2(\varsigma),\ldots, p_m(\varsigma) \right] $. Finally, according to Sturm's Theorem, the number of distinct real roots of $p(x)$ in the half-open interval $(\frac{h_C^2}{2}, R^2]$ is $\sigma(\frac{h_C^2}{2})-\sigma(R^2)$. Sturm's Theorem can help us to quickly determine how many real roots of $p(x)$ are existed in the range $(\frac{h_C^2}{2}, R^2]$ for numerical computation rather than the symbolic computation. \par

\subsection{Algorithm of Optimizing DAE Radius for Path-Loss Exponent 4}
The optimal DAE radius can be calculated by the numerical iterative method as follows. First, use the Sturm's Theorem to determine the number of real roots of (\ref{eight_order_equation}) in the range $(\frac{h_C^2}{2}, R^2)$. Then, find all the real roots of $p(x)$ in the range $(\frac{h_C^2}{2}, R^2)$. Finally, we can get the optimal DAE radius $r^\star$.

\textbf{case 1:} If there is only one real root in $(\frac{h_C^2}{2}, R^2)$, denoted as $x_1$, then we argue that the optimal DAE radius is
\begin{equation}
r^\star=\sqrt{x_1}
\end{equation}

\textbf{case 2:} If there are $\kappa ~(\kappa=2$ or $3$) real roots in $(\frac{h_C^2}{2}, R^2)$, denoted as $\{x_i, i \leq \kappa \}$, the optimal radius is
\begin{equation} \label{opt_r_alpha_4}
r^\star=\arg \max_{r_i, i \leq \kappa} \Upsilon_2(r_i)
\end{equation}
where $r_i=\sqrt{x_i}, i \leq \kappa $. The detailed numerical solving process of the optimal DAE radius $r^\star$ is summarized in Algorithm \ref{alg_1}.
\begin{algorithm}[t]
	\caption{Finding the optimal $r^\star$ using bisection method based on Sturm's Theorem}
	\begin{algorithmic}[1]
		\footnotesize
		\STATE
         Obtain Sturm Sequence $ \left[ p_0(x), p_1(x), p_2(x),\ldots, p_m(x) \right] $ and determine the number of real roots of $p(x)$ in $(\frac{h_C^2}{2}, R^2)$, i.e., $n=\sigma(\frac{h_C^2}{2})-\sigma(R^2)$.\\
         \STATE
         If $n=1$, we get $a_1=\frac{h_C^2}{2}, b_1=R^2$, then skip to step \ref{loop_for} to find the real root $x_1$, thus $r^\star=\sqrt{x_1}$.
         \STATE Else $(n=2$ or $3)$, then isolate the interval $(\frac{h_C^2}{2}, R^2)$ of real roots, resulting in $n$ distinct intervals $(a_1,b_1),\ldots,(a_n,b_n)$, each of which has only one real root and there is no intersection among different intervals. Go to step \ref{loop_for} to find all the real roots $\{x_i, i \leq n \}$ in $(\frac{h_C^2}{2}, R^2)$, thus $r^\star$ is given by (\ref{opt_r_alpha_4}).
         \STATE Endif
        \STATE\label{loop_for}
        For $i=1:1:n$
        \STATE
         Begin\\
         \STATE
         Initialization: $a=a_i, b=b_i$, tolerance $\epsilon >0 $.\\
		\STATE\label{loop_while}
		 While $|a-b| > \epsilon$\\
		\STATE
         Begin\\
         \STATE If $p(\frac{a+b}{2})=0 $, then skip to step \ref{result_x_i}.
         \STATE Elseif $p(a)p(\frac{a+b}{2})>0 $, then $a=\frac{a+b}{2}$.
         \STATE Else, then $b=\frac{a+b}{2}$.
         \STATE Endif
         \STATE End while loop
          \STATE \label{result_x_i}
          $x_i=\frac{a+b}{2}$.
         \STATE End for loop
	\end{algorithmic}
	\label{alg_1}
\end{algorithm}

\section{Simulation Results and Discussion}\label{sec:simulations_discussion}
In this section, we present simulation results and discussion. Specifically, for CA-PB and DA-PB, we give the simulation results of antenna height, average harvested DC power, average WPT efficiency as well as their theoretic values. Parameters used in the simulations are presented in Table I unless stated otherwise.
\begin{table}
\caption{Parameter Setting Used In the Simulation Experiments}
\begin{center}
\begin{tabular}{c c c c}
\hline \hline
      Symbol & Definition & Value  & Unit\\
\hline $h_C$ & Height of CA-PB Antennas & 7.75 &m\\
       $r$   & Radius of UCL Distributed Antennas & 20 & m\\
       $R$   & Radius of Circular Coverage & 30 & m\\
       $I_s$ & Reverse Saturation Current of Schottky Diode &1 & mA\\
       $N$   & Number of Power Beacon Antennas & 100 & \\
       $P$   & Transmit Power of the Power Beacon &20-200 & W\\
       $c$   & Constant Scaling Factor & 1 & \\
       $V_T$ & Thermal Voltage & 28.85 & mV \\
       $\alpha$ & Path-Loss Exponent & 2 or 4 & \\
       $\rho$ & Quality Factor of Schottky Diode & 1 & \\
       $\xi$ & Coefficient of Energy Conversion & 0.85 & \\
       $\sigma_h^2$ & Average Multi-Path Gain & 1 & \\
\hline\hline
\end{tabular}
\end{center}
\end{table}
The transmit power of PB in our simulation experiments is referred to \cite{Survey8}. Note that for the maximal transmit power $P=200$ W and antenna height of CA-PB $h_C=7.75$ m, the maximal radiation power density in the charging cell is 0.265 $\mathrm{W/m^2}$, and much lower than 10 $\mathrm{W/m^2}$, the safety level limited by the FCC. Thus the parameters in our simulation experiments are reasonable.\par
\subsection{Antenna Height of PB}
\begin{figure}[t]
\centering
\includegraphics[width=8cm]{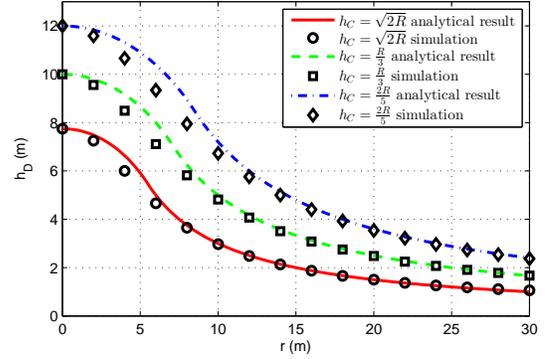}
\caption{The antenna height of DA-PB versus DAE radius $r$, where $N = 100$.}
\label{fig_sim1}
\end{figure}
As is shown in Fig.\ref{fig_sim1}, we illustrate the antenna height of DA-PB when DAE radius becomes larger. Markers in Fig.\ref{fig_sim1} are obtained by exhaustive search of equation (\ref{radiation_CA_equal_to_DA}) while lines are plotted by (\ref{h_D_vs_r}). It is demonstrated that the closed-form result of antenna height is extremely close to the value obtained by exhaustive search of equation (\ref{radiation_CA_equal_to_DA}) for $N=100$ (large scale antenna array). This verifies the closed-form result of antenna height (\ref{h_D_vs_r}). On one hand, the antenna height of DA-PB is a decreasing function of DAE radius; On the other hand, the height of CA-PB in our simulations can make Friis Equation satisfied, i.e., $\frac{h_C^2}{2R}\geq 1$. For the convenience of comparison, we give the results for different antenna heights of CA-PB. It is worth mentioning that lower $h_C$ will surely improve the efficiency of WPT, but it must be elaborately designed with transmit power in order to satisfy safety radiation.\par
\subsection{Average Harvested DC Power}
\begin{figure}[t]
\centering
\includegraphics[width=8cm]{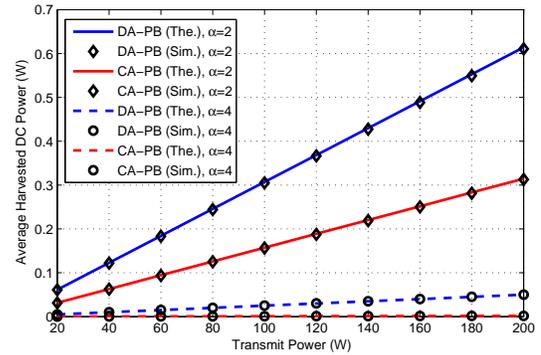}
\caption{The average harvested DC power per user versus transmit power $P$, where $N = 100, r = \frac{2R}{3}$.}
\label{fig_sim2}
\end{figure}
In Fig.\ref{fig_sim2}, we present the simulation results in comparison with the theoretical values. Simulation results are obtained by random realizations of fast fading channel and user locations while theoretical results are obtained by (\ref{ave_P_out_DAS_All}) while $h_D$ is given by exhaustive search of equation (\ref{radiation_CA_equal_to_DA}). It is obvious that simulation results are perfectly consistent with our derived theoretical values. On one hand, it is found that the average harvested DC power for both CA-PB and DA-PB are proportional to the transmit power which can be demonstrated by (\ref{ave_P_out_CAS}) and (\ref{ave_P_out_DAS_All}), respectively; On the other hand, by using DA-PB, the average harvested DC power is larger. \par
\begin{figure}[t]
\centering
\includegraphics[width=8cm]{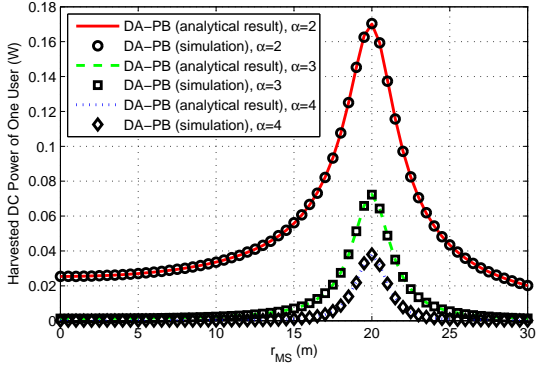}
\caption{The ergodic harvested DC power of one user versus the distance between the user and the cell center $r_{MS}$ for different path-loss exponents $\alpha=2, ~3, ~4$, where $N = 100, r = \frac{2R}{3}, P=20$ W.}
\label{fig_sim3}
\end{figure}
We can see from Fig.\ref{fig_sim3} that the result in (\ref{P_out_DA_r_MS}) when $N \rightarrow \infty$ is extremely consistent with the simulation result when $N$ equals to $100$. Obviously, Fig.\ref{fig_sim3} shows that the ergodic harvested DC power is higher when $r_{MS}$ is close to DAE radius $r$. What's more, for either $r_{MS}>r$ or $r_{MS}<r$, the ergodic harvested DC power is a convex function of $r_{MS}$. As is expected, the smaller path-loss exponent is, the higher ergodic harvested DC power users can harvest.\par
\begin{figure}[t]
\centering
\includegraphics[width=8cm]{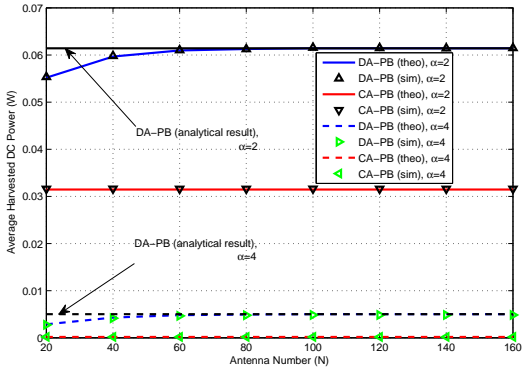}
\caption{The average harvested DC power per user versus antenna number $N$, where $r = \frac{2R}{3}, P=20$ W.}
\label{fig_sim4}
\end{figure}
In Fig.\ref{fig_sim4}, we present the analytical results (i.e., $h_D$ in (\ref{ave_P_out_DAS_All}) are given by (\ref{h_D_vs_r})) as well as simulation results and theoretical values for the average harvested DC power when $N$ becomes larger. Many interesting phenomena can be found from the figure. First, for path-loss exponents $2$ and $4$, the average harvested DC powers by using DA-PB are greater than that by using CA-PB; Second, when the number of DAEs is about $80$, the result we derive under the assumption that $N \rightarrow \infty$ is extremely close to the simulation result, which indicates that the analytical result can be applied in large scale antenna arrays; Finally, the average harvested DC power by using CA-PB is invariant, while the average power harvested by using DA-PB increases when $N$ goes up. This phenomenon shows that by using multi-antennas, the performance gain of our proposed DA-PB can be improved further. In contrast, there is no performance gain when CA-PB uses multiple omnidirectional antennas. \par
\begin{figure}[t]
\centering
\includegraphics[width=8cm]{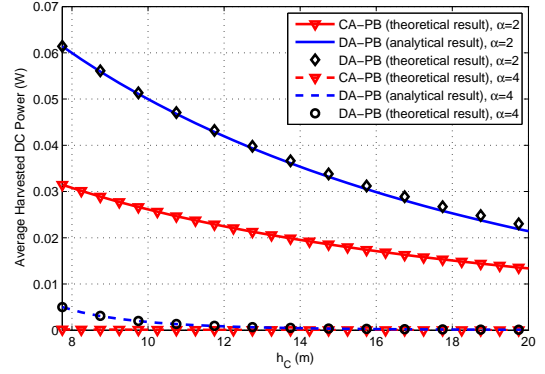}
\caption{The average harvested DC power per user versus antennas height $h_C$, where $N = 100, r = \frac{2R}{3}, P=20$ W.}
\label{fig_sim5}
\end{figure}
As a matter of fact, the antenna height $h_C$ also has an effect on the average harvested DC power. The result in Fig.\ref{fig_sim5} illustrates the effect. Specifically, larger $h_C$ means larger average distances between PB antennas and users, which decreases the average harvested DC power. Even though all the values of average harvested DC power decrease when $h_C$ gradually increases, DA-PB strictly outperforms CA-PB for any arbitrary $h_C$.
\subsection{Average Efficiency of PB}
\begin{figure}[t]
\centering
\includegraphics[width=8cm]{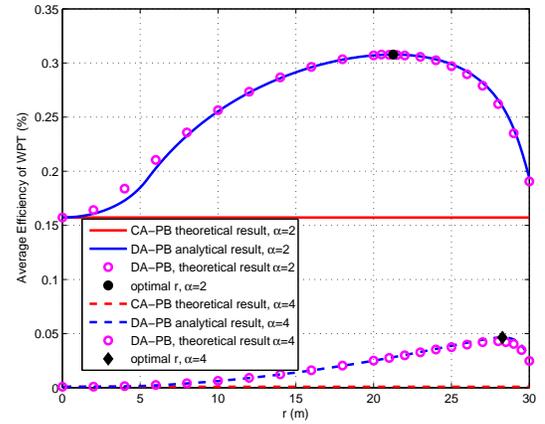}
\caption{The average efficiency of WPT versus DAE radius $r$, where $N=100$.}
\label{fig_sim6}
\end{figure}
In order to verify the optimal DAE radius, we present the simulation results in Fig.\ref{fig_sim6}. Specifically, the magenta hollow circles denote the theoretical values of average efficiency of WPT when antenna number is $100$, while the blue solid line and dashed line denote analytical results for path-loss exponents $2$ and $4$, respectively. For the path-loss exponent $2$, the black solid circle is the optimal DAE radius obtained by (\ref{opt_r_alpha_2}) while for the path-loss exponent $4$, the black solid diamond means the optimal DAE radius using Algorithm \ref{alg_1}. It is obvious that the optimal radii are consistent with the simulation results. Obviously, the DA-PB is strictly better than CA-PB for any DAE radius. Note that the efficiency is lower than one percent, this can be explained as follows. In this paper, in order to satisfy the Friis Equation as well as use simplified path-loss formula, we assume $h_C \geq \sqrt{2R}$. However, $h_C$ could be smaller in practice as long as to restrict the transmit power to satisfy the safety radiation. Thus the average efficiency of WPT could be larger in practice.\par
\begin{figure}[t]
\centering
\includegraphics[width=8cm]{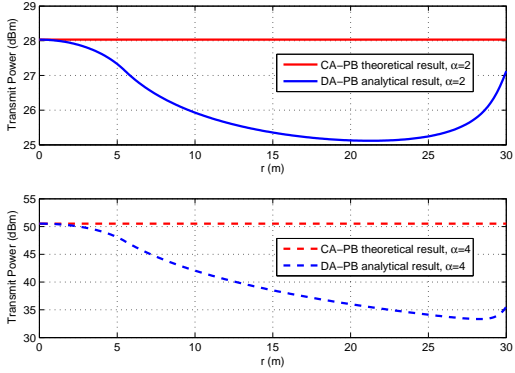}
\caption{The transmit power versus the DAE radius $r$ with average harvested DC power being $1$ mW.}
\label{fig_sim7}
\end{figure}
Compared to CA-PB, DA-PB has other advantages. In Fig.\ref{fig_sim7}, with the average harvested DC power being fixed as 0 dBm (i.e., 1 mW), we find that the transmit power can be dramatically saved by using DA-PB. There is an optimal DAE radius in order to minimize the transmit power. Compared with the case when using CA-PB, for the path-loss exponent $2$, it is easy to find that 3 dB of transmit power can be saved, while more than 15 dB can be saved when path-loss exponent is $4$. This again demonstrates that DA-PB is better than CA-PB.\par
\begin{figure}[t]
\centering
\includegraphics[width=8cm]{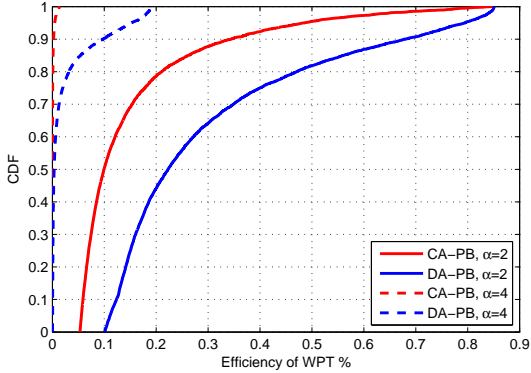}
\caption{The cumulative distribution function of WPT efficiency for users in the charging cell, where $r = \frac{2R}{3}$.}
\label{fig_sim8}
\end{figure}

As we can see from Fig.\ref{fig_sim8}, the cumulative distribution function (CDF) of WPT efficiency of CA-PB is significantly steeper than that of DA-PB for path-loss exponent $2$ and $4$. This indicates that there is a larger area that users can harvest more power by using DA-PB than that by using CA-PB. The efficiency of CA-PB is extremely lower compared with DA-PB. For example, when path-loss exponent is $2$, the probabilities of efficiency being larger than $0.5$ percent are $0.2$ for DA-PB and $0.05$ for CA-PB, respectively. This phenomenon can be explained as follows. First, CA-PB with longer average propagation distance means higher propagation path-loss which reduces the WPT efficiency; Second, by using DA-PB, the average distance between DAEs and users is shortened, which decreases the path-loss of the power transfer and eventually increases the WPT efficiency. Note that the WPT efficiency can be further improved by lowing $h_C$ so long as to restrict the transmit power to satisfy the safety radiation.\par

\section{Conclusion}\label{sec:conclusion}
In this paper, we consider a novel antenna deployment of PB, i.e., DA-PB. We derive the antenna height of DA-PB to protect users from being hurt by RF electromagnetic radiation. Besides, we get the average harvested DC power per user in the charging cell. In order to maximize the average efficiency of DA-PB, we get the optimal DAE radius of circularly distributed PB antennas. Finally, simulation results verify the theoretical results and show that the proposed DA-PB indeed achieves larger average harvested DC power per user and average efficiency of WPT than conventional CA-PB. These useful observations can give operators valuable directions when exploiting PBs in WPT or future Wireless Powered Communications Network (WPCN).
\section*{Acknowledgment}
The work is supported by Natural Science Basic Research Plan in Shaanxi Province of China (Program No.2015JQ6234) and the National Natural Science Foundation of China (NSFC) under Grant (No.61461136001).

\newcounter{TempEqCnt}                         
\setcounter{TempEqCnt}{\value{equation}} 
\setcounter{equation}{41}
\begin{figure*}[ht]
\begin{small}
\begin{equation}\label{Q_alpha_2}
\begin{split}
Q&=2\int_{0}^{R}\underbrace{\int_{0}^{\pi}\frac{\rho}{(\rho^2+r^2+h_D^2-2\rho r\cos \theta)}\ud\theta}_{I} \ud\rho=\int_{0}^{R}\frac{2 \pi \rho}{\sqrt{(\rho^2+r^2+h_D^2)^2-4\rho^2 r^2}}\ud \rho \\
& \xlongequal{t=\rho^2}\int_{0}^{R^2}\frac{\pi}{\sqrt{(t^2+2(h_D^2-r^2)t+(r^2+h_D^2)^2}} \ud t = \pi \left[\operatorname{arcsinh}\left(\frac{R^2+h_D^2-r^2}{2rh_D}\right)-\operatorname{arcsinh}\left(\frac{h_D^2-r^2}{2rh_D}\right)\right]
\end{split}
\end{equation}
\end{small}
\hrulefill
\end{figure*}
\setcounter{equation}{\value{TempEqCnt}}

\newcounter{TempEqCnt2}                         
\setcounter{TempEqCnt2}{\value{equation}} 
\setcounter{equation}{45}
\begin{figure*}[ht]
\begin{small}
\begin{equation}\label{h11}
\begin{split}
&h_{11}(r)\!=\!2r \left[ \sqrt{(R^2\!+\!h_C^2\!-\!2r^2)^2\!+\!\frac{(2h_C^2\!-\!4r^2)^2}{h_C^2\!-\!r^2}}\left(R^2\!-\!h_C^2\right) +R^2\left(R^2-2r^2\right)+\left(h_C^2-2r^2\right)^2+\frac{(2h_C^2-4r^2)^2}{h_C^2-r^2} \right]
\end{split}
\end{equation}
\end{small}
\hrulefill
\end{figure*}
\setcounter{equation}{\value{TempEqCnt2}}

\newcounter{TempEqCnt3}                         
\setcounter{TempEqCnt3}{\value{equation}} 
\setcounter{equation}{46}
\begin{figure*}[ht]
\begin{small}
\begin{equation}\label{h12}
\begin{split}
h_{12}(r) &= \frac{h_C^4}{32r^7} \left[\left(4r^2R^2\!+\!h_C^4\!-\!4r^4\right)^2+16r^4h_C^4 + \left(4r^2R^2\!+\!h_C^4\!-\!4r^4\right) \sqrt{\left(4r^2R^2\!+\!h_C^4\!-\!4r^4\right)^2+16r^4h_C^4} \right.\\
&~~~~ - \left. \left(h_C^4\!+\!4r^4\right) \sqrt{\left(4r^2R^2\!+\!h_C^4\!-\!4r^4\right)^2+16r^4h_C^4} - \left(h_C^4\!+\!4r^4\right) \left(4r^2R^2\!+\!h_C^4\!-\!4r^4+4r^2h_C^4\right) \right] \\
\end{split}
\end{equation}
\end{small}
\hrulefill
\end{figure*}
\setcounter{equation}{\value{TempEqCnt3}}

\begin{appendices}
\section{  }
The radiation power density at the coordinate of $(\nu,0,0)$ is
\begin{equation}
\begin{split}
\Psi_d(\nu)&\!=\!\sum\limits_{i=1}^N\! \frac{\frac{P}{4\pi N}}{\left(\nu\!-\!r\cos\frac{2\pi(i-1)}{N}\right)^2\!+\!\left(r\sin\frac{2\pi(i-1)}{N}\right)^2\!+\!h_D^2} \\
&= \frac{P}{4\pi N} \frac{N}{2\pi}\sum\limits_{i=1}^N \frac{1}{r^2+\nu^2-2r\nu \cos\frac{2\pi(i-1)}{N}+h_D^2} \frac{2\pi}{N} \\
\end{split}
\end{equation}
so
\begin{equation}\label{nu_star_cal}
\nu^\star=\arg\max_{\nu \in [0,R]}{\Psi_d(\nu)}
\end{equation}
It's very hard to get $\nu^\star$ from (\ref{nu_star_cal}). So we can not give a closed-form expression of $h_D$ for arbitrary $N$ from (\ref{radiation_CA_equal_to_DA}). For $N\rightarrow\infty$, the radiation power density is
\begin{equation}\label{ays_Psi_d}
\begin{split}
\lim_{N\rightarrow\infty}{\Psi_d}(\nu) &= \frac{P}{8\pi^2}\int_{0}^{2\pi} \frac{1}{r^2+\nu^2-2r\nu \cos\theta +h_D^2} \ud\theta \\
&= \frac{P}{4\pi} \frac{1}{\sqrt{(r^2+\nu^2+h_D^2)^2-4r^2\nu^2}}
\end{split}
\end{equation}
where (\cite{Survey20}, (3.661.4)) was exploited to derive (\ref{ays_Psi_d}). Thus, for $N\rightarrow\infty$, (\ref{nu_star_cal}) is equivalent to
\begin{equation}
\nu^\star=\arg\min_{\nu \in [0,R]} ~{\left(r^2+\nu^2+h_D^2\right)^2-4r^2\nu^2}
\end{equation}
Let $t=\nu^2$, we have $f(t)=t^2+2(h_D^2-r^2)t+(r^2+h_D^2)^2$. With $f'(r)=0$, we get
\begin{equation}
t^\star= r^2-h_D^2
\end{equation}

\textbf{case 1:} If $t^\star>0$, we argue that $\nu^\star=\sqrt{t^\star}=\sqrt{r^2-h_D^2}$, thus the maximal radiation power density is $\frac{P}{8 \pi r h_D}$. According to (\ref{radiation_CA_equal_to_DA}), we have
\begin{equation}
h_D=\frac{h_C^2}{2r}
\end{equation}

\textbf{case 2:} If $t^\star \leq 0$, we argue that $\nu^\star=0$. Similarly to Case 1, we get
\begin{equation}
h_D=\sqrt{h_C^2-r^2}
\end{equation}
From the above, we conclude
\begin{equation}
\begin{split}
&{\nu^\star}=
\begin{cases}
0, & \mbox~0\leq r \leq \frac{h_C}{\sqrt{2}},\\
\sqrt{r^2-\left(\frac{h_C^2}{2r}\right)^2}, & \mbox ~\frac{h_C}{\sqrt{2}}\leq r \leq R.\\
\end{cases}
\end{split}
\end{equation}
and
\begin{equation}
\begin{split}
&{h_D}=
\begin{cases}
\sqrt{h_C^2-r^2}, & \mbox~0\leq r \leq \frac{h_C}{\sqrt{2}},\\
\frac{h_C^2}{2r}, & \mbox ~\frac{h_C}{\sqrt{2}}\leq r \leq R.\\
\end{cases}
\end{split}
\end{equation}
Thus this ends the proof.

\section{  }
It is difficult to give a closed-form expression of $Q$ for arbitrary path-loss exponent $\alpha$, but we get a closed-form result when $\alpha$ takes the typical value $2$ and $4$. Specifically, for the special case $\alpha=2$, $Q$ is derived as (\ref{Q_alpha_2}) where (\cite{Survey20}, (3.661.4)) and (\cite{Survey20}, (2.261)) were exploited to derive $I$ and the last equation in (\ref{Q_alpha_2}), respectively. With $\operatorname{arcsinh}(x)=\ln\left(x+\sqrt{x^2+1}\right)$, and after some algebraic manipulations, we finally get
\setcounter{equation}{42}
\begin{equation}
Q=\pi \ln\left(\frac{R^2+h_D^2-r^2+\sqrt{(R^2+h_D^2-r^2)^2+4r^2h_D^2}}{2h_D^2}\right)
\end{equation}
For $\alpha=4$, similar derivation procedure can be followed to get $Q$. Thus this ends the proof.

\newcounter{TempEqCnt4}                         
\setcounter{TempEqCnt4}{\value{equation}} 
\setcounter{equation}{53}
\begin{figure*}[ht]
\begin{small}
\begin{equation}\label{h21}
\begin{split}
h_{21}(r) &= 2r \left[ \left(R^2+h_C^2\right)\sqrt{R^4+R^2(2h_C^2-4r^2)+h_C^4} + \left(R^2+h_C^2\right)\left(h_C^2-r^2\right)\frac{2R^2}{\sqrt{R^4+R^2(2h_C^2-4r^2)+h_C^4}} \right.\\
&~~ \left. +2R^2\left(h_C^2-r^2\right) +R^2\left(R^2-2r^2\right) +h_C^4 \right]
\end{split}
\end{equation}
\end{small}
\hrulefill
\end{figure*}
\setcounter{equation}{\value{TempEqCnt4}}

\newcounter{TempEqCnt5}                         
\setcounter{TempEqCnt5}{\value{equation}} 
\setcounter{equation}{54}
\begin{figure*}[ht]
\begin{small}
\begin{equation}\label{h22}
\begin{split}
h_{22}(r) &= \! A \left\{\frac{h_C^4}{2r^3}\left[R^4+R^2\left(\frac{h_C^4}{2r^2}-2r^2\right)+\left(r^2+\frac{h_C^4}{4r^2}\right)^2\right]^{\frac{3}{2}} +\frac{4r^2R^2h_C^4\!-\!8r^4h_C^4}{8r^5}\left[R^4\!+\!R^2\left(\frac{h_C^4}{2r^2}\!-\!2r^2\right)\!+\!\left(r^2+\frac{h_C^4}{4r^2}\right)^2\right] \right.\\
&~~~~ -\left.\frac{h_C^4(4r^2R^2-h_C^4-4r^4)}{32r^4}\left[R^2\left(-\frac{h_C^4}{r^3}-4r\right) + 2\left(r^2+\frac{h_C^4}{4r^2}\right)\left(2r-\frac{h_C^4}{2r^3}\right)\right] \right\}
\end{split}
\end{equation}
\end{small}
\hrulefill
\end{figure*}
\setcounter{equation}{\value{TempEqCnt5}}

\section{  }
For the special case $\alpha=2$, the optimization problem $\mathbf P1$ can be reduced to the following problem
\begin{equation}
\begin{split}
\mathbf~~~~&\max_{r }~f_1(r)\\
&~\mathrm{s.t.}~~ 0 \leq r\leq {R}
\end{split}
\end{equation}
where $f_1(r)=\ln\left(\frac{R^2+h_D^2-r^2+\sqrt{(R^2+h_D^2-r^2)^2+4r^2h_D^2}}{2h_D^2}\right)$ and $h_D$ is given by (\ref{h_D_vs_r}). For the convenience of calculation, let $a \triangleq R^2+h_D^2-r^2$,$b \triangleq h_D^2$,$c \triangleq 2rh_D$, thus the first-order derivative of $f_1(r)$ is given by
\begin{equation}
f_1'(r)\!=\!\frac{\left(a'\sqrt{a^2+c^2}\!+\!aa'\!+\!cc'\right)b\!-\!\left(a\sqrt{a^2+c^2}\!+\!a^2\!+\!c^2\right)b'}{\left(a+\sqrt{a^2+c^2}\right)b \sqrt{a^2+c^2}}
\end{equation}
With the nominator being always larger than zero, we only consider the numerator.

\textbf{case 1:} When $r \in \left(0,\frac{h_C}{\sqrt{2}}\right)$, denote the numerator as (\ref{h11}).
For any $h_C \in [\sqrt{2R}, R)$, it is easy to show that $h_{11}(r)>0$ always holds. Therefore, for $r \in \left(0,\frac{h_C}{\sqrt{2}}\right)$, $f_1'(r)>0$ always holds. Note that there is a minimal value of $f_1(r)$ when $r=0$, so we discard it and only focus on $r>0$ from now on.

\textbf{case 2:} When $r \in \left(\frac{h_C}{\sqrt{2}}, R\right)$, denote the numerator as (\ref{h12}).
Discarding the positive terms and after some algebraic manipulations, we get
\setcounter{equation}{47}
\begin{equation}
\begin{split}
I_1(r)\!&=\!4r^2 \left[\left(R^2\!-\!2r^2\right)\left(\sqrt{(4r^2R^2\!+\!h_C^4\!-\!4r^4)^2\!+\!16r^4h_C^4} \right.\right.\\
&~~ \left.\left. +4r^2R^2+h_C^4-4r^4\right)+4r^2h_C^4 \right]
\end{split}
\end{equation}
With the variable substitution $x=r^2$, let $I_1(x)=0$. We get
\begin{equation}
4x^2-2R^2x-h_C^4=0
\end{equation}
Note that $x$ is larger than zero, so
\begin{equation}
x_1=\frac{R^2 + \sqrt{R^4+4h_C^4}}{4}
\end{equation}
For any $h_C \in [\sqrt{2R}, R)$, it is easy to show that $x_1 \in \left(\frac{h_C^2}{2}, R^2\right)$. Therefore, the uniqueness of root of equation $f_1'(r)=0$ in the range $\left(\frac{h_C}{\sqrt{2}}, R\right)$ has been demonstrated. It is easy to show that $f_1'(r)\mid_{r \rightarrow \frac{h_C}{\sqrt{2}}^{-}} = f_1'(r)\mid_{r \rightarrow \frac{h_C}{\sqrt{2}}^{+}} > 0$, so the $f_1'(r)$ is continuous at $r=\frac{h_C}{\sqrt{2}}$. On one hand, with $f_1'(r)>0$ for $r \in \left(0,\frac{h_C}{\sqrt{2}}\right]$, which has been proved above, we argue that the optimal DAE radius must lie in the range $\left(\frac{h_C}{\sqrt{2}}, R\right]$. On the other hand, $f_1'(r)\mid_{r \rightarrow R^-}<0 $, $f_1(R)$ is certainly not the maximal value. Therefore
\begin{equation}
   r^\star=\sqrt{x_1}=\frac{1}{2}\sqrt{R^2+\sqrt{R^4+4h_C^4}}
\end{equation}
This ends the proof.

\section{  }
For the special case $\alpha=4$, similar derivation procedure can be followed to get the optimal DAE radius. The optimization problem $\mathbf P2$ can be reduced to the following problem
\begin{equation}
\begin{split}
\mathbf~~~~&\max_{r }~f_2(r)\\
&~\mathrm{s.t.}~~ 0 \leq r\leq {R}
\end{split}
\end{equation}
where $f_2(r)=\frac{R^2-h_D^2-r^2+\sqrt{R^4+R^2(2h_D^2-2r^2)+(r^2+h_D^2)^2}}{h_D^2 \sqrt{R^4+R^2(2h_D^2-2r^2)+(r^2+h_D^2)^2}}$ and $h_D$ is given by (\ref{h_D_vs_r}). Let $a \triangleq h_D^2$,$b \triangleq \sqrt{R^4+R^2(2h_D^2-2r^2)+(r^2+h_D^2)^2}$,$c \triangleq R^2-h_D^2-r^2$, thus the first-order derivative of $f_2(r)$ is given by
\begin{equation}
f_2'(r)=\frac{\left(c'+b'\right)ab-\left(c+b\right)\left(a'b+ab'\right)}{\left(ab\right)^2}
\end{equation}

\textbf{case 1:} When $r \in \left(0,\frac{h_C}{\sqrt{2}}\right)$, denote the numerator as (\ref{h21}).
Similar to $\alpha=2$, for $r \in \left(0,\frac{h_C}{\sqrt{2}}\right)$, it is easy to prove that $f_2'(r)>0$ always holds. So we discard it and only focus on $r>0$ from now on.

\textbf{case 2:} When $r \in \left(\frac{h_C}{\sqrt{2}}, R\right)$, denote the numerator as (\ref{h22}),
where $A=\left[R^4\!+\!R^2\left(\frac{h_C^4}{2r^2}\!-\!2r^2\right)\!+\!\left(r^2\!+\!\frac{h_C^4}{4r^2}\right)^2\right]^{-\frac{1}{2}}$. Discarding the positive terms and after some algebraic manipulations, we get
\setcounter{equation}{55}
\begin{equation}
\begin{split}
&I_2(r) =\left[16r^4R^4+8r^2R^2(h_C^4-4r^4)+(4r^4+h_C^4)^2\right]^{\frac{3}{2}} \\
&\!+\!\left[16r^4R^4\!+\!8r^2R^2(h_C^4\!-\!4r^4)\!+\!(4r^4\!+\!h_C^4)^2\right]\left(4r^2R^2\!-\!8r^4\right) \\
&\!-\! \left(4r^2R^2\!-\!h_C^4\!-\!4r^4\right)\left[-4r^2R^2(h_C^4\!+\!4r^4)\!+\!16r^8\!-\!h_C^8\right]
\end{split}
\end{equation}
With the variable substitution $x=r^2$, let $I_2(x)=0$. We get
\begin{equation}
\begin{split}
   &256x^8\!-\!768R^2x^7\!+\!128(\!6R^4\!+\!h_C^4\!)x^6\!+\!(\!224h_C^4R^2\!-\!256R^6\!)x^5 \\
   &-192R^4h_C^4x^4\!-\!32R^2h_C^4(\!R^4+2h_C^4\!)x^3\!-\!8h_C^8(\!4R^4+h_C^4\!)x^2 \\
   &-10R^2h_C^{12}x-h_C^{16}=0
\end{split}
\end{equation}
Similar to $\alpha=2$, it can be proved that the optimal real root must lie in the range $(\frac{h_C^2}{2}, R^2)$. Therefore, the optimal DAE radius must be one of the square-roots of the above eight-order equation real roots in $(\frac{h_C^2}{2}, R^2)$. This ends the proof.
\end{appendices}

\end{document}